\documentclass[%
reprint,
amsmath,amssymb,
prb,
floatfix,
]{revtex4-2}

\usepackage[utf8]{inputenc}

\usepackage{graphicx}
\usepackage{dcolumn}
\usepackage{bm}

\usepackage{booktabs}

\usepackage{multirow}

\usepackage{siunitx}

\usepackage{xcolor}
\colorlet{linkBlue}{blue!30!black}

\usepackage{hyperref}
\hypersetup{pdfpagemode={UseOutlines},
bookmarksopen=true,
bookmarksopenlevel=0,
hypertexnames=false,
colorlinks=true,
citecolor=linkBlue,
linkcolor=linkBlue,
urlcolor=linkBlue,
pdfstartview={Fit},
unicode=true,
breaklinks=false,
}

\begin{document}

\title{Ferroelectric, quantum paraelectric or paraelectric? Calculating the evolution from BaTiO$_3$ to SrTiO$_3$ to KTaO$_3$ using a single-particle quantum-mechanical description of the ions}
\author{Tobias Esswein}
\email{tobias.esswein@mat.ethz.ch}
\author{Nicola A. Spaldin}
\email{nicola.spaldin@mat.ethz.ch}
\affiliation{Materials Theory, Department of Materials, ETH Zurich, Switzerland}
\date{\today}

\begin{abstract}
We present an inexpensive first-principles approach for describing quantum paraelectricity that combines density functional theory (DFT) treatment of the electronic subsystem with quantum mechanical treatment of the ions through solution of the single-particle Schrödinger equation with the DFT-calculated potential. Using BaTiO$_3$, SrTiO$_3$ and KTaO$_3$ as model systems, we show that the approach can straightforwardly distinguish between ferroelectric, paraelectric and quantum paraelectric materials, based on simple quantities extracted from standard density functional and density functional perturbation theories. We calculate the influence of isotope substitution and strain on the quantum paraelectric behavior, and find that, while complete replacement of oxygen-16 by oxygen-18 has a surprisingly small effect, experimentally accessible strains can induce large changes. Finally, we collect the various choices for the phonon mass that have been introduced in the literature. We identify those that are most physically meaningful by comparing with our results that avoid such a choice through the use of mass-weighted coordinates.
\end{abstract}

\maketitle

\section{Introduction}\label{sec:intro}

Quantum paraelectric materials, such as strontium titanate, SrTiO$_3$ (STO) and potassium tantalate, KTaO$_3$ (KTO), are incipient ferroelectrics whose ferroelectric phase transition on cooling is suppressed by quantum fluctuations~\cite{MullerSrTiO31979}. They are characterized by a temperature-dependent transverse optical phonon, whose frequency tends to but does not reach zero at low temperature~\cite{Cowley:1962}, and a correspondingly high low-temperature dielectric susceptibility~\cite{WeaverDielectric1959} that deviates from classical Curie behavior with a cross-over to a characteristic $T^{-2}$ scaling below a few tens of kelvin~\cite{FujishitaQuantum2016,RowleyFerroelectric2014}. Their proximity to ferroelectricity means that the ferroelectric state can be reached readily with external perturbations, including pressure~\cite{UweStressinduced1976,FujiiStressInduced1987}, homovalent A-site~\cite{BednorzSr1984} or B-site substitution~\cite{HochliQuantum1977,RytzDielectric1980}, oxygen isotope substitution~\cite{ItohFerroelectricity1999,ItohQuantum2000} and strain~\cite{HaeniRoomtemperature2004,TyuninaEvidence2010}. This sensitivity, combined with the low temperatures, makes detailed characterization of the structure of the quantum paraelectric state challenging: Neutron and x-ray Rietveld analysis of STO, for example, indicate a centrosymmetric structure down to \SI{1.5}{K}~\cite{KiatRietveld1996}. In contrast, nuclear magnetic resonance (NMR) shows local dynamic polar off-centering of the Ti ions~\cite{ZalarNMR2005}, consistent with the anomalous vibrational amplitudes seen in $\gamma$-ray Bragg scattering~\cite{JauchAnomalous1999}, and scanning transmission electron microscopy reveals local polar nanoregions~\cite{Salmani-RezaiePolar2020}. 

\begin{figure}[!h]
  \includegraphics[width=\linewidth, keepaspectratio]{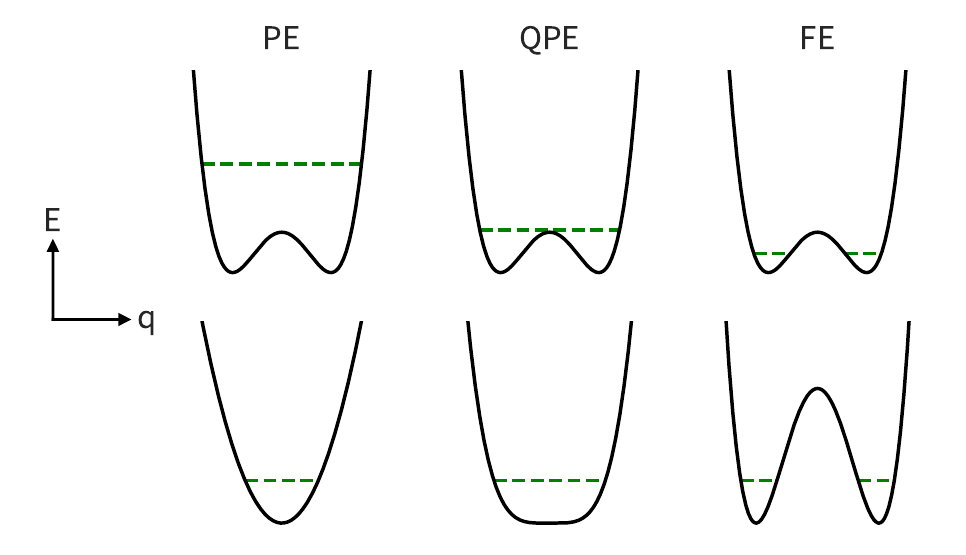}
  \caption{\label{fig:DWintro}
  Illustration of two possible mechanisms for the crossover from the paraelectric (left) to ferroelectric (right) state. In each panel the vertical axis is internal energy, and the horizontal axis is the polarization (or relative displacements of the anionic and cationic sublattices) in mass-weighted coordinates $q$, with the curves centered around zero polarization. The green dashed lines represent the zero-point energy levels. The lower row shows a behavior analogous to the usual Landau model of displacive phase transitions as a function of temperature: The internal energy has a double well potential resulting in a ferroelectric state at low temperature (right panel) with the barrier height reducing as temperature increases, so that there is a single minimum corresponding to zero polarization above the Curie temperature. Such a crossover could also occur at zero kelvin as a function of an external parameter such as strain or pressure, in which case the middle panel, in which the potential has a pronounced quartic component but not yet a barrier, would correspond to the quantum paraelectric state. In the upper row, the free energy is unchanged as a function of tuning parameter, but the zero-point energy evolves from high (corresponding to the paraelectric state) to low (corresponding to the ferroelectric state). In this case, quantum paraelectric behavior would be expected when the zero-point energy is in the vicinity of the top of the barrier between the oppositely polarized states. This behavior could be achieved by changing the masses of the atoms through isotopic substitution, which changes the mass without changing the form of the potential energy. 
  }
\end{figure}

From a theoretical perspective, the quantum paraelectric behavior is broadly understood to result from quantum fluctuations suppressing the softening of the polar phonon, which would otherwise drive a ferroelectric phase transition at low temperature. Indeed, extension of the classic Slater model of ferroelectricity~\cite{SlaterLorentz1950} to treat the soft polar mode quantum mechanically~\cite{BarrettDielectric1952,KhmelnitskiiLowtemperature1971} and with anharmonic coupling to other phonon modes modes~\cite{MullerSrTiO31979,BendikModel1972}, correctly reproduces the $T^{-2}$ scaling of the dielectric susceptibility. The importance of a shallow ``double well'' potential energy surface with its correspondingly small zero-point energy (Fig.~\ref{fig:DWintro}) has been emphasized~\cite{TosattiRotational1994}, with the isotope effect then explained by suppression of the zero-point motion by larger atomic masses~\cite{KvyatkovskiiTheory2001}. First-principles electronic structure calculations based on density functional theory (DFT) confirm the shallow double-well picture~\cite{ShinQuantum2021a} and indicate the importance of transition-metal -- oxygen polarizability~\cite{MigoniOrigin1976,BilzTheory1987}, manifesting in anomalously large Born effective charges and giant LO-TO splittings~\cite{King-SmithFirstprinciples1994,ZhongGiant1994}. DFT studies have also explored the relationship between polar distortions and tetragonality, strain, unit-cell volume and oxygen octahedral rotations~\cite{SaiFirstprinciples2000,AschauerCompetition2014}, as well as the effect of point defects on the structural properties~\cite{LuoStructural2004}. 

Particularly important insights are provided by atomistic simulations using path-integral quantum Monte Carlo methods, in which the ionic motions are treated quantum mechanically. In such simulations, quantum effects are found to suppress the ferroelectric transition in STO completely as well as to strongly affect the behavior of the polar mode up to $\sim$\SI{100}{K}; in KTO, quantum effects cause local correlated polar nanoregions~\cite{AkbarzadehAtomistic2004}. In contrast, ionic quantum effects have a minimal effect in the prototypical ferroelectric barium titanate, BaTiO$_3$ (BTO)~\cite{ZhongEffect1996}, with its high ferroelectric ordering temperature. Unfortunately, however, path-integral quantum Monte Carlo methods are computationally expensive even at high temperatures, and become unfeasible on approaching zero kelvin, and so are impractical for routine evaluation of material properties. 

While many features of quantum paraelectricity are now established, major questions about the detailed nature of the quantum paraelectric state remain open. First, the existence or not of a zero-kelvin quantum critical point. While susceptibility measurements have been interpreted in terms of a quantum critical point that is crossed by either strain or $^{18}$O substitution~\cite{RowleyFerroelectric2014,CoakDielectric2018,CoakQuantum2020}, the zero-kelvin limit has not in fact been reached~\cite{RytzDielectric1980}, and instead an up-turn in dielectric susceptibility has been reported at very low temperatures~\cite{RowleyFerromagnetic2010,RowleyFerroelectric2014}. The up-turn points to a discontinuous transition, which is indeed captured by a Hamiltonian based on a polarizability model~\cite{Bussmann-HolderDimensional2008}, and has been attributed to coupling to strain~\cite{CoakDielectric2018,CoakQuantum2020}. Second, and related to the previous point, is the appropriateness of a double-well picture for describing the crossover from paraelectric through quantum paraelectric to ferroelectric behavior. Finally, unconventional superconductivity has been observed in both STO~\cite{SchooleySuperconductivity1964} and KTO~\cite{UenoDiscovery2011,ChenTwoDimensional2021,LiuTwodimensional2021}, with ferroelectric quantum fluctuations proposed as the source of the pairing mechanism~\cite{Edge_et_al:2015}. While the results of subsequent experiments, particularly the effects of strain and $^{18}$O substitution, have proved consistent with the predictions of the quantum fluctuations model~\cite{StuckyIsotope2016,RuhmanSuperconductivity2016,CoakPressure2019,CollignonMetallicity2019,vanderMarelPossible2019,GastiasoroSuperconductivity2020}, further insight into the nature of the quantum criticality would be invaluable in better understanding and further optimizing the superconductivity.

In this work we use electronic structure calculations based on standard density functional theory (DFT), to explore the potential energy landscapes of a series of quantum paraelectrics and conventional ferroelectrics. Since conventional DFT does not treat the ions quantum mechanically, we then account for the quantum mechanical behavior of the ions in a second step by solving the Schrödinger equation explicitly for the ionic subspace using the calculated DFT potential. Our approach is similar to that used in Ref.~\cite{ShinQuantum2021a}, with the important difference that we work in mass-weighted coordinates and so are not forced to make  arbitrary assumptions about the sizes of the phonon mass and displacement. 

Our primary scientific goal is to answer the question of how a ferroelectric emerges from a paraelectric as a function of an external (non-thermal) tuning parameter. In Fig.~\ref{fig:DWintro} we sketch two commonly discussed scenarios for this crossover, each of which might be applicable in certain regimes. In all cases the vertical axis is internal energy and the horizontal axis is the soft-mode coordinate $q$. The lower panel shows the usual Landau theory picture in which a paraelectric (PE) parabolic potential evolves into a ferroelectric (FE) double-well potential via a strongly anharmonic single-well potential, with the zero-point energy remaining largely constant. In this scenario, which might be achieved for example by applying pressure or strain, the quantum paraelectric (QPE) regime corresponds to the intermediate strongly anharmonic, but still single-well, potential. The upper panel of Fig.~\ref{fig:DWintro} shows a complementary limit in which the double-well potential is unchanged across the transition but the zero-point energy shifts. Such behavior could be induced by isotope substitution, with the zero-point energy shifting down as the ionic masses are increased, and QPE would be expected in the region for which the zero-point energy coincides with the top of the double well. A second, methodological goal is to explore whether simple, inexpensive standard DFT-based methods are appropriate for addressing questions that are explicitly related to quantum tunneling of ions. Such a finding would, in turn, allow easy determination of whether new hypothetical materials could exhibit quantum paraelectric behavior.

\section{Methods \& Theoretical Approach}\label{sec:methods}

In this section we describe our methodology for calculating the potential wells of the type shown in Fig.~\ref{fig:DWintro}, and solving the resulting Schrödinger equations for these potentials. 

\subsection{Mass-weighted coordinates and the Schrödinger equation}
At first sight it is trivial to solve the 1D Schrödinger equation, 
\begin{equation}
    \left( -\dfrac{\hbar^2}{2m}\dfrac{d^2}{d\xi^2} + V(\xi) \right) \psi(\xi) = E \psi(\xi) ,    
\end{equation}
given the calculated form of the potential, $V$, as a function of ionic displacements, $\xi$, and knowing the masses of the ions, $m$, contributing to the phonon eigenvector. In practice, however, for the lattice vibrations in periodic solids considered here, there is a subtlety, in that neither the {\it mass} nor the {\it displacement} of a phonon is well-defined. Many plausible choices have been made in the literature and we return to this point in sections \ref{sec:masses} and \ref{sec:displacements}, where we provide a summary of the literature choices use our results to determine which are most appropriate. Here, we prefer to combine the mass and displacement into mass-weighted coordinates, $q$, defined as $ q = \sqrt{m}\xi $, which are rigorously defined and avoid an arbitrary choice for the phonon mass and displacement. In mass-weighted coordinates, the 1D Schrödinger equation is reformulated as 
\begin{equation}
    \left( -\dfrac{\hbar^2}{2}\dfrac{d^2}{dq^2} + V(q) \right) \psi(q) = E \psi(q) \quad ,   
\end{equation}
which we then solve using a finite-difference approach to obtain the zero-point and higher energy levels. 
\subsection{Construction of the potential V(q)}\label{sec:potconstruct}

\begin{figure}[h!]
  \includegraphics[width=\linewidth, keepaspectratio]{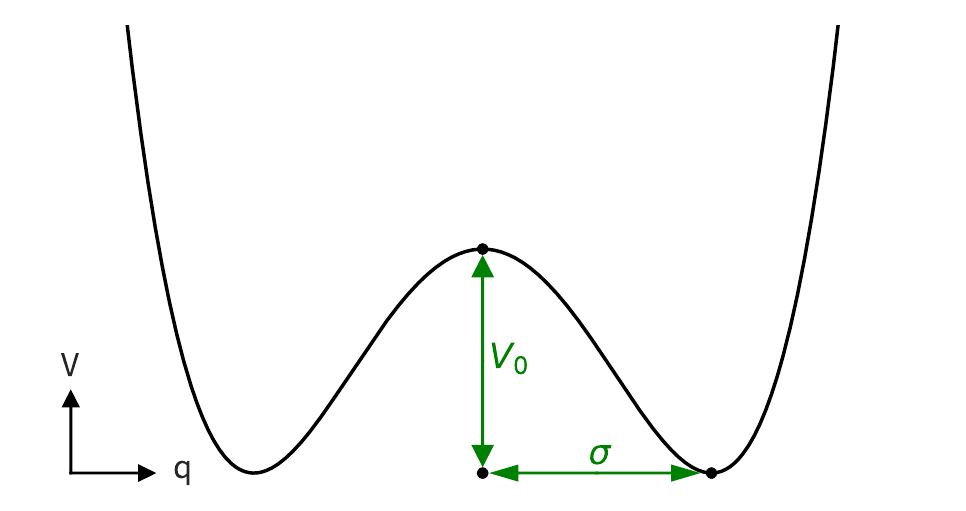}
  \caption{\label{fig:DWconstruction} 
  Sketch of the double-well construction in mass-weighted coordinates, indicating the parameters $V_0$ and $\sigma$. The height of the double well barrier, $V_0$, is calculated from the energy difference between the polar and non-polar structures. The half-width of the double well, $\sigma$, (defined here as the distance between the barrier top and the bottom on one side) is calculated from the imaginary phonon frequency of the non-polar structure at the $\Gamma$-point, $\omega$, and the energy difference $V_0$ as described in the text. 
  }
\end{figure}

We write the double-well potential $V(q)$ as a fourth-order polynomial 
\begin{equation}
    V(q) = V_0 (\dfrac{q^4}{\sigma^4} - 2 \dfrac{q^2}{\sigma^2} + 1) \quad,
    \label{Eqn:Vofq}
\end{equation}
where $V_0$ is the height of the barrier and $\sigma$ is the half-width of one well, as shown in Fig.~\ref{fig:DWconstruction}. The ``$+1$'' sets the zero of energy to the bottom of the well rather than the top of the barrier. We find that this simple fourth-order polynomial deviates only slightly from a full calculation of the energy as a function of polar distortion over the relevant energy range.

The height of the barrier, $V_0$, is extracted straightforwardly from the energy difference per formula unit between the non-polar reference structure and the lower-energy polar structure. 
The half-width of the barrier, $\sigma$, is obtained from the frequency, $\omega$ of the imaginary phonon at the $\Gamma$ point as follows: By definition, for a harmonic potential $\omega^2$ is related to the curvature by
\begin{equation}
    \omega^2 = \dfrac{d^2 V(q)}{dq^2}\bigg\rvert_{q=0} \quad.
    \label{Eqn:w2}
\end{equation}
(Note that in this case $\omega^2$ is negative). Taking the second derivative of Eqn.~\ref{Eqn:Vofq} and setting $q=0$ yields
\begin{equation}
    \dfrac{d^2 V(q)}{dq^2}\bigg\rvert_{q=0} = - \dfrac{4 V_0}{\sigma^2} \quad ,
    \label{Eqn:d2Vdq2}
\end{equation}
and equating Eqns.~\ref{Eqn:w2} and \ref{Eqn:d2Vdq2} we obtain
\begin{equation}
    \sigma^2 = -\dfrac{4 V_0}{\omega^2}  \quad .    
\end{equation}

We mention that this one-dimensional model corresponds physically to the case where the system inverts its polarization via the high-symmetry zero-polarization reference structure, rather than for example rotating the polarization into another orientation. As such, it provides an upper bound on the barrier height between the oppositely polarized states.

\subsection{Computational details}

To calculate the forces and total energies needed to construct our potentials, we use density functional theory within the generalized gradient approximation (GGA) as implemented in the Quantum Espresso~6.4.1 code \cite{GiannozziQUANTUM2009,GiannozziAdvanced2017}. 
We describe the exchange-correlation functional using the PBE functional~\cite{PerdewGeneralized1996}, and perform the core-valence separation with the GBRV pseudopotentials~\cite{GarrityPseudopotentials2014,GarrityGBRV2019}. We use a kinetic energy cutoff of \SI{60}{Ry} (\SI{816}{eV}) for the wavefunctions, and a $\Gamma$-centered 16$\times$16$\times$16~k-point mesh for all unit cells. 
Total energies are converged to \SI{1}{\micro eV} (\SI{7.35e-8}{Ry}) and forces to \SI{0.1}{meV/\AA} (\SI{3.89e-6}{Ry/Bohr}). 

In a first step, we calculate the lattice constants and internal coordinates of a non-polar reference structure for each material as follows: For BTO we fully relax the atomic positions and lattice parameters to obtain the known experimental low-temperature rhombohedral phase with polarization along the pseudocubic [111] direction; we then remove the polar distortion by hand while keeping the lattice parameters fixed. For STO, we construct a non-polar tetragonal unit cell containing the experimentally observed rotations of the oxygen octahedra around the [001] direction, then relax the atomic positions and cell parameters while constraining the symmetry to maintain the inversion center. For KTO we relax the lattice constants for the primitive cubic unit cell with the symmetry constraint that the atoms remain at their high-symmetry cubic perovskite positions and the unit cell remains cubic. 
We then calculate the ferroelectric soft-mode phonon frequencies at $\Gamma$ for these non-polar reference structures using density functional perturbation theory (DFPT), as implemented in Quantum Espresso~6.4.1. In all cases these frequencies are imaginary (that is $\omega^2$ is negative), indicating a double-well potential.

In a second step, we calculate the polar structures by manually adding a polarization to each material following the eigenvector of the imaginary phonon mode, then relaxing the internal coordinates, while keeping the shape and volume of the unit cells fixed. For BTO, the resulting structure is the original polar structure of the first step. The energy difference between each polar and corresponding non-polar structure gives us directly the $V_0$ parameter in each case. This is then combined with the calculated $\omega^2$ value to obtain the well half-width, $\sigma$.

\section{Double wells, zero-point energies and tunneling frequencies}\label{sec:results}

Next, we present and analyze our calculated potential energy double wells, as well as the zero-point and higher energies obtained from solving the respective Schrödinger equations, for our example materials BTO, STO, and KTO, as well as for experimentally plausible generalized double wells. 

\subsection{Trends across the BaTiO$_3$, SrTiO$_3$, KTaO$_3$ series}

\begin{figure*}[ht]
  \includegraphics[width=\linewidth, keepaspectratio]{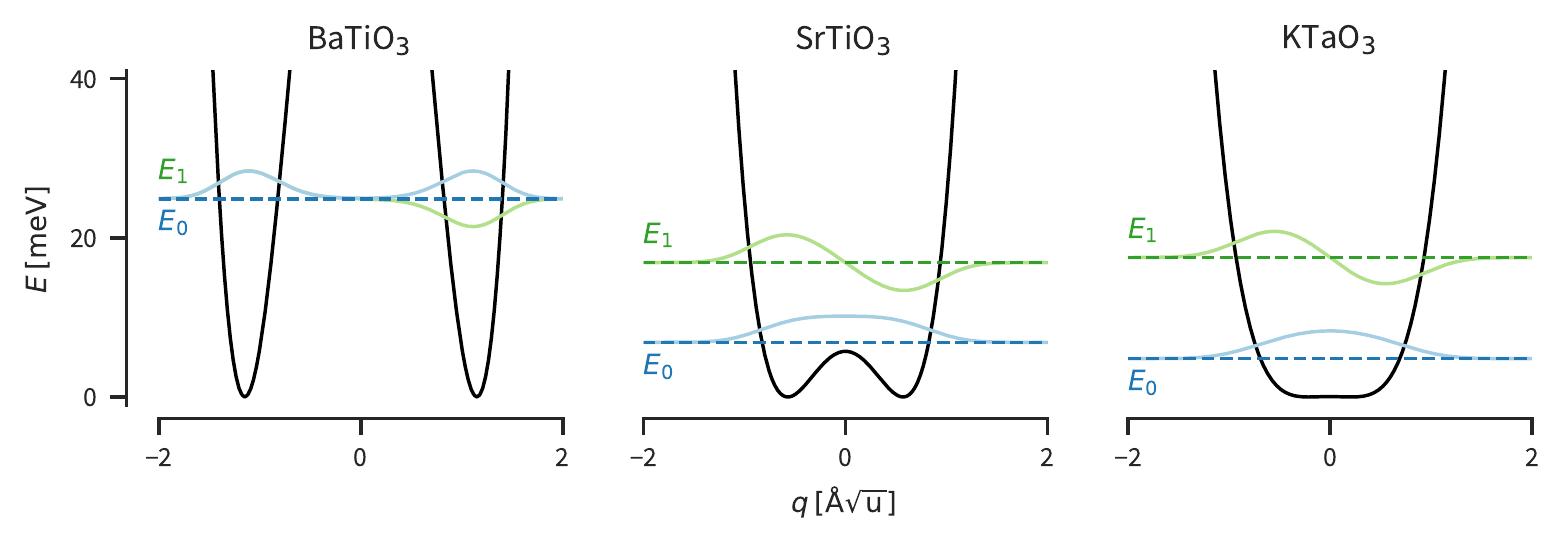}
  \caption{\label{fig:results_BSK}
  Ferroelectric double-well potentials of BTO, STO and KTO calculated in this work using  density functional theory (solid black lines), and their respective zero-point (dashed blue) and first excited state (dashed green) energies obtained by subsequent solution of the Schrödinger equation. We note that the trend follows that of the lower panel of  Fig.~\ref{fig:DWintro}. In ferroelectric BTO (left), the zero-point energy and first excited state are deep in the double-well and almost degenerate. In quantum paraelectric STO (middle), the zero-point energy is around the same height as the double-well barrier.  In KTO (right), the zero-point energy is higher than the double-well barrier, indicating that KTO, while still having quantum paraelectric features, is closer to conventional paraelectric behavior than STO. Corresponding numerical data are given in Tab.~\ref{tab:DWvalues}. 
  }
\end{figure*}

In Fig.~\ref{fig:results_BSK} we show our calculated potential energies (black solid lines), as well as the two lowest-energy eigenvalues (blue and green dashed lines) for BaTiO$_3$, SrTiO$_3$, and KTaO$_3$. The energy scale is the same for all materials, and each horizontal axis is in mass-weighted coordinates.

We note first that all materials have a double-well potential, with both the height and width of the barrier decreasing from ferroelectric BTO to paraelectric KTO. The barrier height of BTO is $\sim$\SI{105}{meV}, which is high compared to $k_BT$ at room temperature, consistent with its ferroelectric ground state; likewise the lowest energy eigenvalue (dashed blue line) lies far below the top of the barrier, and its eigenfunction (solid blue line) is largely localized in one or other of the oppositely polarized wells. The energy of the first excited state (dashed green line) is indistinguishable from the zero-point energy at this scale. This small splitting suggests a low tunneling frequency for a system initially localized in one energy well to tunnel to the other, again consistent with the known ferroelectric ground state of BTO. Extracting a numerical ``tunneling frequency'', $f$, from $f=\frac{E_1 - E_0}{h}$ (see Table~\ref{tab:DWvalues}) gives a value in the low-GHz range, much lower than the frequencies for STO and KTO. We do not, of course, expect the ferroelectricity in BTO to switch at this ``tunneling frequency'', since it does not capture the physics of domain wall motion, but we find it a useful quantity for comparing trends across our series of materials.  

The barrier height of STO is more than one order of magnitude smaller than that of BTO, and its width in mass-weighted coordinates is approximately half. The zero-point energy lies close to the top of the barrier, and the ground-state eigenfunction has its maximum at $q=0$; this corresponds to the ``extreme quantum'' regime of Ref.~\cite{TosattiRotational1994}. The splitting between the ground and first excited states is \SI{10.02}{meV}, leading to a tunneling frequency of \SI{2.4}{THz}, intriguingly close to the measured \SI{5}{K} soft-mode phonon frequency of \SI{0.5}{THz}~\cite{YamadaNeutron1969,VogtRefined1995,YamanakaEvidence2000}. 

While KTO still has a double-well potential for our choice of computational parameters, the barrier is small. The zero-point energy is clearly above the barrier, indicating that KTO should behave more like a conventional paraelectric than STO. The splitting between the ground and first excited states is \SI{12.71}{meV} corresponding to $\frac{E_1 - E_0}{h} =  3.07$ THz. In this case, it is more appropriate to describe this value as an oscillation rather than a tunneling frequency, since the relevant energy levels are above the barrier.

\begin{table}[h!]
    \caption{\label{tab:DWvalues}%
    Calculated double-well barriers, $V_0$, and half-widths, $\sigma$, and resulting energy splittings, $E_1 - E_0$, and tunneling frequencies, $f$, of BTO, STO and KTO, corresponding to the double wells shown in Figs.~\ref{fig:results_BSK} (upper three rows of the table) and \ref{fig:results_iso_vol} (lower four rows of the table).
    }
    \begin{tabular}{l r r r r}
    \toprule
           & \multicolumn{2}{c}{barrier}  &   \multicolumn{2}{c}{tunneling} \\
    \cmidrule(lr){2-3}    \cmidrule(lr){4-5}
           & $\sigma$ [\si{\AA \sqrt{u}}] & $V_0$ [\si{meV}] & $E_1-E_0$ [\si{meV}] & $f$ [\si{THz}] \\
    \midrule
        BaTiO$_3$          &  1.149 &  104.8   &   0.008 &  0.002  \\
        SrTiO$_3$          &  0.571 &    5.74  &  10.02  &  2.42   \\
        KTaO$_3$           &  0.215 &    0.056 &  12.71  &  3.07   \\
    \midrule
        SrTi$^{18}$O$_3$     &  0.591 &    5.74  &   9.38  &  2.27  \\
        KTa$^{18}$O$_3$      &  0.223 &    0.056 &  12.09  &  2.92  \\
        \midrule
        SrTiO$_3$  V.+1\% &  0.646 &    9.52  &   7.95  &  1.92  \\
        KTaO$_3$  V.+1\%  &  0.471 &    1.35  &  10.60  &  2.56  \\
    \bottomrule
    \end{tabular}
\end{table}

\subsection{Isotope effect}

\begin{figure*}[ht]
  \includegraphics[width=\linewidth, keepaspectratio]{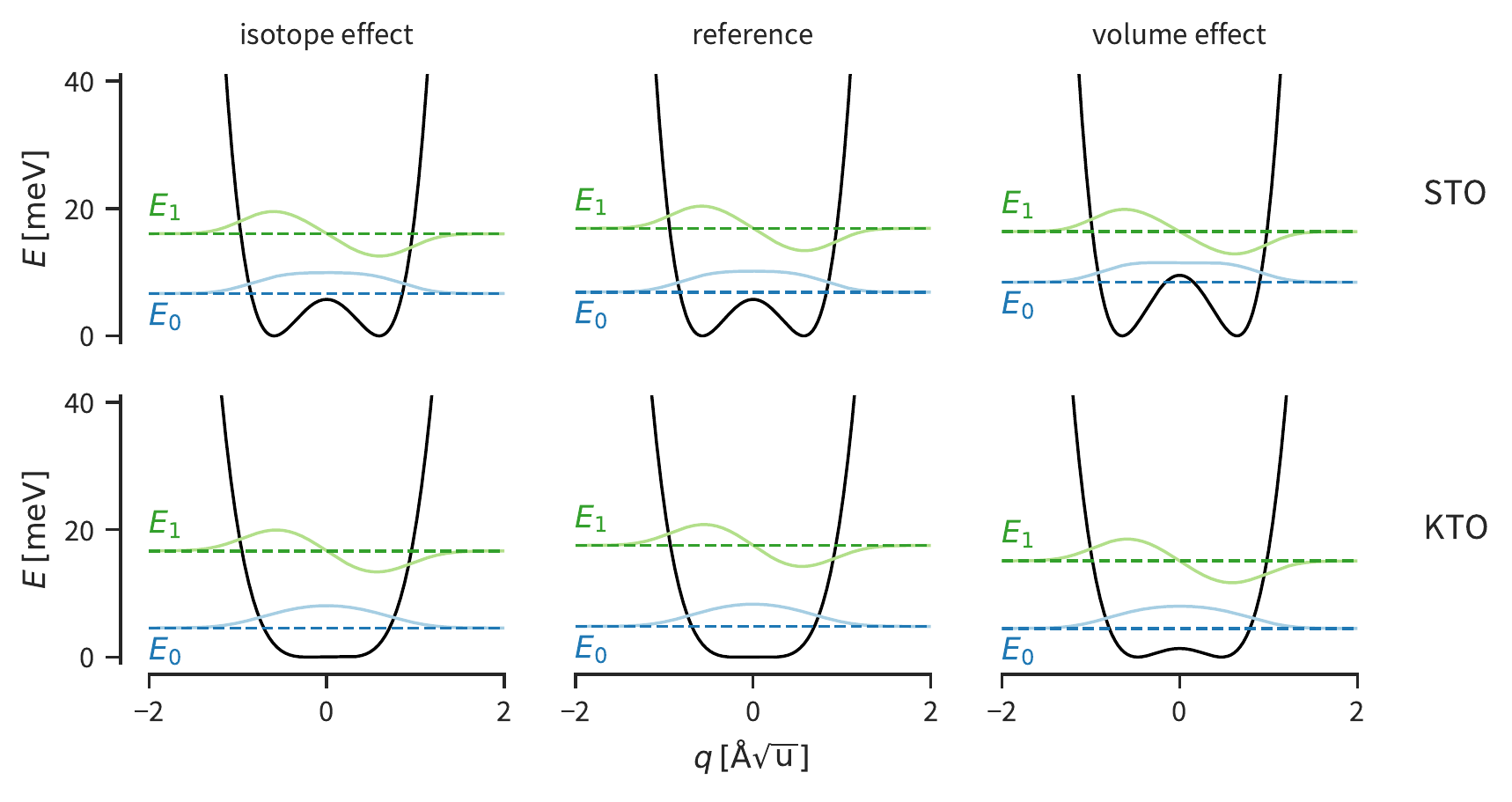}
  \caption{\label{fig:results_iso_vol}
  Double wells and lowest-energy eigenvalues and wavefunctions for $^{18}$O-substituted (left column) and \SI{1}{\%}-volume-expanded STO (top) and KTO (bottom). The middle column shows the reference calculations from Fig.~\ref{fig:results_BSK} for comparison. Numerical data are given in Tab.~\ref{tab:DWvalues}. We see that heavy-isotope substitution widens the wells, while keeping the barrier heights constant, resulting in a lowering of the zero-point energies and energy differences. A \SI{1}{\%} increase in volume changes the well-widths and energies in the same directions by a slightly larger amount. 
  }
\end{figure*}

Our calculated double wells and low-energy eigenvalues and wavefunctions in $^{18}$O-substituted STO (top row) and KTO (bottom row) are shown in the left column of Fig.~\ref{fig:results_iso_vol}; the reference data from Fig.~\ref{fig:results_BSK} are shown in the central column. For both STO and KTO we see that the heavy-oxygen double well is slightly wider than the reference one, since the heavier oxygen ions result in a lower phonon frequency, $\omega$. This in turn causes lower zero-point energy levels, smaller energy differences between the lowest energy levels, and oscillation frequencies (\SI{2.27}{THz} for STO and  \SI{2.92}{THz} for KTO) that are $\sim 5$~{\%} smaller than in the reference cases. These small changes are surprising, in light of the strong changes in ferroelectric and superconducting properties reported for $^{18}$O-substituted STO in particular~\cite{StuckyIsotope2016,RuhmanSuperconductivity2016,CoakPressure2019,CollignonMetallicity2019,vanderMarelPossible2019,GastiasoroSuperconductivity2020}. 

\subsection{Effect of strain}

Changes in lattice constant, introduced by applying uniaxial stress, or through biaxial coherent heteroepitaxy, have also been shown to have a substantial effect on both ferroelectric and superconducting behaviors of SrTiO$_3$~\cite{Haeni_et_al:2004,HerreraStrainengineered2019}. We investigate the effect of strain on the double-well profiles and properties next. In the right column of Fig.~\ref{fig:results_iso_vol} we show our calculated double wells and energy eigenvalues and eigenfunctions for STO (top) and KTO (bottom) when the volume of the unit cell is increased by \SI{1}{\%} (without isotope substitution). This corresponds to pseudocubic lattice constants of $a=b=\SI{3.9422}{\AA}$ and $c=\SI{3.9613}{\AA}$ for STO and $a=b=c=\SI{4.0268}{\AA}$ for KTO, which is an increase of slightly more than \SI{0.013}{\AA} for each direction. Note that volume changes of this order have been reported in transition-metal oxides on changing the point defect chemistry by annealing in reducing or oxidizing atmospheres\cite{GrandeAnisotropic2012}. We find that a volume increase causes changes in the same direction as an increase in isotopic mass, with a 1\% volume increase having a markedly stronger effect than complete $^{16}$O~$\rightarrow$~$^{18}$O substitution. Compared to the reference structure, the barrier height in STO is almost doubled, and the lowest energy eigenvalue lies just below the barrier, indicating that STO moves into the ferroelectric regime under strain. In KTO, the barrier height increases by a factor of 24, and both energy eigenvalues move lower and closer together. 

An increase in {\it lattice constant} (rather than volume) of 1\%, which is accessible in biaxially strained coherent thin-film heterostructures, causes even larger changes (not shown in Figs.~\ref{fig:results_iso_vol} and \ref{fig:results_2d}): In STO, the energy barrier (\SI{20.54}{meV}) cuts through the lowest two energy eigenvalues and, combined with the $\sigma$ of \SI{0.765}{\AA \sqrt{u}}, the tunneling frequency is lowered to \SI{1.01}{THz},  less than half of the reference frequency. In KTO, the changes are even more significant, with a crossover from the oscillating to the tunneling regime occurring. The double well width and height are \SI{0.757}{\AA \sqrt{u}} and \SI{10.07}{meV}, giving a tunneling frequency of \SI{1.34}{THz}, reduced from the reference \SI{3.07}{THz}. 

Since the lattice constants of complex oxides are strongly sensitive to oxygen stoichiometry~\cite{GrandeAnisotropic2012}, it is clear that care must be taken in comparing the quantum paraelectric behavior of different STO and KTO samples. In particular, if the process of isotopic substitution also changes the oxygen vacancy concentration, it might be difficult to disentangle intrinsic isotopic effects from changes in behavior resulting from changes in the lattice constant. 

\subsection{Tunneling / oscillation frequencies for general double-well potentials}

\begin{figure}[h!]
  \includegraphics[width=\linewidth, keepaspectratio]{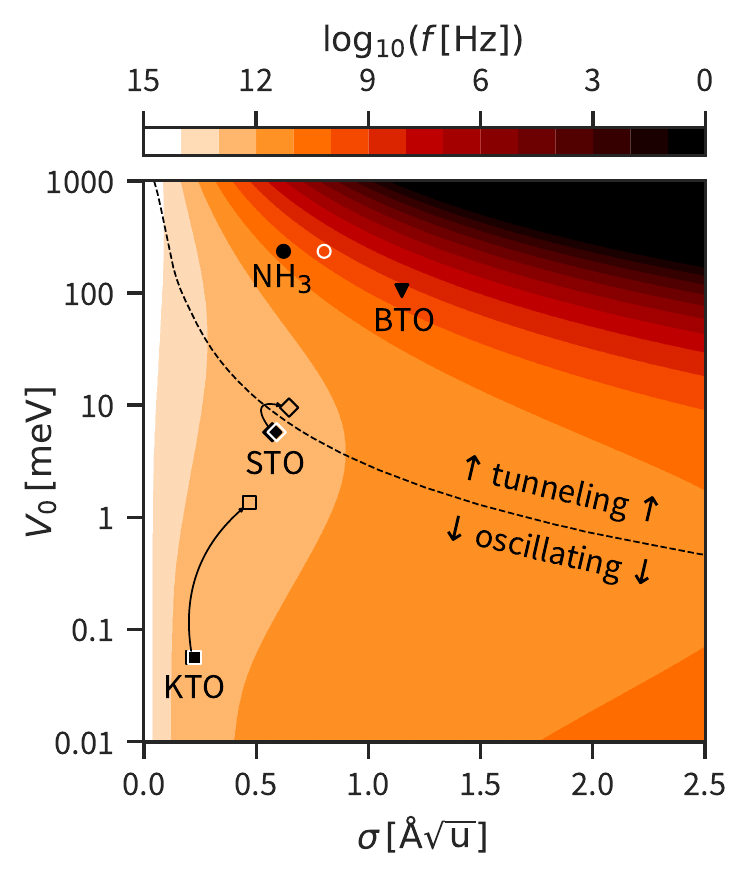}%
  \caption{\label{fig:results_2d}
  2D map of our calculated oscillation/tunneling frequencies (color bar) for quartic double-well potentials with barrier height $V_0$ ranging from 0.01 to 1000 meV (note the log scale) and barrier half-width, $\sigma$ less than 2.5 \AA$\sqrt{u}$. The black dashed line separates the  ``oscillating'' region (lower left), in which the zero-point energy is above the double-well barrier, from the ``tunneling'' region (upper right), in which the zero-point energy is below than the double-well barrier. Frequencies are color-coded on a logarithmic scale, ranging from white on the left (corresponding to frequencies \SI{>100}{THz}), through orange (THz and GHz) and red (MHz and kHz) to black (\SI{<10}{Hz}). Solid black symbols show the frequency for each material calculated in this work. White open symbols show the frequencies for the corresponding isotope-substituted materials; black open symbols, the frequencies at 1\% increased volumes. 
  }
\end{figure}

In Fig.~\ref{fig:results_2d} we present a map of the tunneling frequencies, $f$, calculated using our approach, for a physically relevant range of barrier heights, $V_0$, between 0.01 and \SI{1000}{meV}, and barrier half-widths, $\sigma$, between 0.01 and \SI{2.5}{\AA \sqrt{u}}. Our motivation is to provide a convenient chart for looking up the tunneling or oscillation frequency, and hence the proximity to quantum paraelectric behavior, for any material, given the double-well height and width. Frequencies, ranging from high-THz (white) on the left, to low-Hz (black) on the top right, are color-coded from white to black, and the crossover from ``tunneling'' (zero-point energy $<$ double-well barrier) to ``oscillating'' (zero-point energy $>$ double-well barrier) is shown by the dashed black line (top left to center right). 

Before analyzing our example materials, we first discuss some general trends. A narrow free energy barrier (left border) generally results in high oscillation frequencies, and the narrower the well, the higher the barrier required to reach the tunneling regime. Widening the well (by increasing $\sigma$) lowers the frequency, which can be reduced to the high-GHZ range, even for the smallest barrier heights, $V_0$, shown. At any given barrier half-width $\sigma$, increasing the barrier height from close to zero first increases the frequency, until the barrier height reaches the zero-point energy; for larger barrier heights  the frequency decreases with increasing barrier height. The point where the zero-point energy crosses the double-well barrier, indicated by the dashed line in the plot, therefore corresponds to the frequency maximum and marks the crossover between the oscillating and tunneling regimes. We propose that this crossover provides the best measure of the quantum critical point within the density functional theory formalism. 

The solid black symbols show the results of our calculations for BTO, STO, and KTO presented earlier, as well as for an isolated molecule of ammonia, NH$_3$, using $V_0$ and $\sigma$ values directly taken from ref.~\cite{HalpernIntrinsic2001}. The white outlined circle, diamond and square show the results of our calculations for NH$_3$, STO, and KTO with D substituted for H in the NH$_3$ case, and $^{18}$O substituted for $^{16}$O in STO and KTO. The minimal effect of 100\% oxygen isotope substitution pointed out earlier is strikingly clear in this visualization. In the case of ammonia, in contrast, the isotope substitution of deuterium for hydrogen moves the zero-point energy substantially lower in the double well, and lowers the tunneling frequency correspondingly, consistent with experimental observations. Our calculated results for STO and KTO with 1~\% volume increase are shown with the black outlined diamond and square, connected to the reference values by black arrows. In both cases the substantial effect mentioned above is clearly visible: For STO, the volume effect leads to a crossover from the oscillating to the tunneling regime; for KTO, there is a strong shift towards tunneling behavior. We emphasize again that even a small volume change during the isotope substitution process will likely overshadow any direct isotope effect, and one has to take special care to exclude this factor when attributing changes in behavior to changes in the masses of the atoms.

\section{Appropriate choices of phonon effective mass and displacement}\label{sec:massdisp}

In this final section, we collect the many definitions that have been used in the literature for the mass of the phonon. We compare the predictions of the various choices with the results that we obtained above using mass-weighted coordinates, for which no arbitrary choice has to be made. Our main finding is that one has to take care when separating mass and displacement to obtain physically sensible results. In particular, the popular choice of the B-site ion off-centering as a measure of the polar displacement, while giving accidentally reasonable values for the titanates, is not generally physically appropriate. For the perovskite oxides, the combination of the mass of the oxygen ions in the unit cell for the effective mass, combined with the average displacement of an oxygen ion between para- and ferro-electric structures as the effective displacement is the most self-consistent choice.

\subsection{Phonon effective mass}\label{sec:masses}

In Table~\ref{tab:masses} we collect the mass values for BTO, STO and KTO for the various definitions that have been used in the literature. All numbers are in units of atomic mass, $u$.

\begin{table*}[htb]
    \caption{\label{tab:masses}%
    Values of the BTO, STO and KTO phonon effective masses, $m^*$ (in $u$) for various definitions used in the literature (top three rows). The bottom three rows show the displacement value (in \AA) for each mass definition, calculated according to $\delta = \sigma / \sqrt{m*}$, using the $\sigma$ values that we reported in Table~\ref{tab:DWvalues}. 
    }
    \begin{tabular}{c l r r r r r r r r}
    \toprule
        & & \multicolumn{6}{c}{based only on masses} & \multicolumn{2}{c}{including eigenvector info.} \\
    \cmidrule(lr){3-8} \cmidrule(lr){9-10}
        & & ABO$_3$ & B & O$_3$ & reduced & AB/O$_3$ & Wentzcovich & eigenvector & curvature \\
        & &   & \cite{BarrettDielectric1952} & \cite{NakamuraPerovskiterelated1997} & \cite{AtkinsMolecular2011} & \cite{PalOrigin2021} & \cite{ShinQuantum2021a} & \cite{ZhongGiant1994,ZhongEffect1996} &  \\
    \midrule
        \multirow{3}*{\rotatebox[origin=c]{90}{masses}\hspace{
        0mm}} \multirow{3}*{\rotatebox[origin=c]{90}{($u$) }\hspace{2mm}}%
        & BaTiO$_3$ &   233.2 &    48 &    48 &     4.6 &  38 &  18 &    22 &  0.022  \\
        & SrTiO$_3$ &   183.5 &    48 &    48 &     4.6 &  35 &  14 &    22 &  0.036  \\
        & KTaO$_3$  &   268.0 &   181 &    48 &     4.6 &  39 &  20 &    22 &  0.005  \\
    \midrule
        \multirow{3}*{\rotatebox[origin=c]{90}{displace-}\hspace{-1mm}}
        \multirow{3}*{\rotatebox[origin=c]{90}{ment (\AA)}\hspace{2mm}}%
        & BaTiO$_3$ & 0.075 & 0.166 & 0.166 & 0.536 & 0.186 & 0.271 & 0.245 &  3.65   \\
        & SrTiO$_3$ & 0.042 & 0.082 & 0.082 & 0.266 & 0.097 & 0.153 & 0.122 &  3.01   \\
        & KTaO$_3$  & 0.013 & 0.016 & 0.031 & 0.100 & 0.034 & 0.048 & 0.046 &  3.04   \\
    \bottomrule
    \end{tabular}
\end{table*}

\subsubsection{Based only on masses}

The definitions in the first six columns in Table~\ref{tab:masses} use only the masses of the constituent atoms. 

Column 1 shows the ABO$_3$ \textit{atomic} mass, which is just the sum of the individual atoms in one formula unit. STO is the lightest at \SI{183.5}{u}, followed by BTO and KTO with \SI{233.2}{u} and \SI{268.0}{u}, respectively. A problem with this definition is that the effective mass is the same for each phonon mode; this is clearly unphysical since different phonons correspond to different patterns of atomic displacements. For example, the heavy barium atom in BTO and tantalum atom in KTO contribute strongly to their respective atomic masses, but do not displace much in the soft mode, which is dominated by motion of the lighter atoms.

A popular choice in the ferroelectrics community is the mass of the B-site atom (column 2), that is titanium (\SI{47.9}{u}) for BTO, and STO~\cite{BarrettDielectric1952}, and tantalum (\SI{180.9}{u}) for KTO. This choice is motivated by the fact that ferroelectric polarization results in large part from the relative displacement of the B-site cation and its oxygen coordination octahedron. We note, however, that the A-site cation is also displaced from its high-symmetry position in the ferroelectric ground state, and that the cations contribute less to the dynamical displacements in the soft mode than the oxygen anions due to their large mass. 

With this in mind, the masses of the lighter oxygens (column 3) probably provide a better definition for the mass of the soft mode in perovskite oxides, since they have the largest displacements in the polar phonons. Their combined mass is coincidentally equal to that of titanium, as already observed by \citeauthor{NakamuraPerovskiterelated1997}~\cite{NakamuraPerovskiterelated1997}, possibly explaining why choice of the Ti mass yields sensible results in perovskite-oxide ferroelectric titanates.

The generalized, mode-independent reduced masses~\cite{AtkinsMolecular2011}, defined as $\mu = (\sum_{i} \frac{1}{m_i})^{-1}$ (column 4) are \SI{4.6}{u} for all three materials. Since the reduced mass is dominated by the lightest atom, the value will be similar for all perovskite oxides. A recently proposed modified reduced mass~\cite{PalOrigin2021}, $m_{eff} = (\frac{1}{m_{A}+m_{B}} + \frac{1}{m_{O_3}})^{-1}$, (column 5) is also dominated by the oxygen ions and gives almost the same value for all three materials considered.

Recently, use of a ``Wentzcovich-type'' mass, defined as $\frac{3*M_{tot}}{4\pi^2}$ and first introduced in the context of molecular-dynamics simulations with variable cell shape~\cite{WentzcovitchInvariant1991} was proposed in the ferroelectrics context~\cite{ShinQuantum2021a}. Our values using this formula are shown in column 6 and obviously are smaller than the ABO$_3$ masses but with the same relative values. 

\subsubsection{Incorporating information about the eigenvectors}

The definitions shown in columns 7 and 8 incorporate information about the phonon eigenvector as well as the masses of the atoms. 

In the PI-QMC simulations of refs.~\cite{ZhongGiant1994,ZhongEffect1996}, the pseudoparticle mass was calculated as the sum of each atomic mass times its displacement in the normalized phonon mode eigenvector squared. Interestingly, this eigenvector mass, listed in column 7 for the softest phonon mode in each case, has almost the same value in all three materials, and is almost independent (varying by $<5\%$) of the particular low-energy polar phonon mode at $\Gamma$ chosen for the calculation. 

Finally, a phonon effective mass can be defined by analogy to the electronic effective mass, from the quadratic part of the phonon dispersion around the $\Gamma$ point, using $m^* = \frac{\hbar^2}{\partial^2 E / \partial k^2}$. (Note that this can not be used for the acoustic modes, which are linear at $\Gamma$). We list our calculated values in the final column, noting that they are much smaller than those obtained from all other methods, and also that the precise values are strongly sensitive to the detailed numerics of the phonon calculation. 

\subsubsection{From experiment}

Finally, we mention that an effective mass was extracted from experimental measurements of the domain-wall motion near the ferroelectric quantum critical point in TTF-QBr$_2$I$_2$, using a simple Wentzel-Kramers-Brillouin approximation~\cite{KagawaAthermal2016}. The value extracted in this way is roughly half the proton mass, which is three orders of magnitude smaller than the mass of the TTF-QBr$_2$I$_2$ molecule. 

\subsubsection{Summary of phonon effective mass definitions}

The preceding discussion clearly illustrates the difficulty with assigning a mass to a phonon, with the physically justifiable values given in Table~\ref{tab:masses}  spanning a range of several orders of magnitudes, from less than \SI{1}{u} to \SI{268}{u} for the materials considered here. In addition, most of the approaches reviewed assign a mass per formula unit or unit cell, without considering interactions or coupling between unit cells. Since polar nanometer-sized domains have recently been observed in strained STO thin films close to the quantum critical point~\cite{Salmani-RezaieOrderDisorder2020,Salmani-RezaiePolar2020}, an appropriate correction could be to multiply any chosen effective mass $m_{\mathrm{eff}}$ by a factor corresponding to the number of correlated unit cells~\cite{BrookeTunable2001}. Such an argument was used 
in ref.~\cite{ZhongEffect1996} to justify a large effective mass, and correspondingly minimal quantum effects, for the antiferrodistortive rotational mode of STO. 

\subsection{Effective displacements}\label{sec:displacements}

The same argument that precludes an unambiguous definition of phonon mass also results in the phonon displacement not being unambiguously defined. Our results obtained with mass-weighted coordinates, however, allow us to extract the appropriate displacement within the double-well potential model corresponding to a particular choice of the mass. As reported in Table~\ref{tab:DWvalues}, our calculated values of $\sigma$ are \SI{1.15}{\AA \sqrt{u}}, \SI{0.57}{\AA \sqrt{u}} and \SI{0.22}{\AA \sqrt{u}} for BTO, STO and KTO respectively. In the lower rows of Table~\ref{tab:masses} we list the calculated displacements, $\delta = \sigma / \sqrt{m*}$, resulting from these $\sigma$ values for each choice of mass. 

For comparison, in Table~\ref{tab:displacements} we list (in \AA) the calculated displacement of each individual atom between the high-symmetry centrosymmetric and low-symmetry polar phases, corrected for any shift in the center of mass. For the O$_3$ and ABO$_3$ columns, the value is the average absolute displacement of all ions, with the direction of displacements of the oxygens being opposite to that of the cations. As expected, the displacement of the lightest oxygen ions is largest in all three materials, although the displacements of the Ti ions in BTO and STO are also substantial, consistent with the usual Slater model of the ferroelectric soft mode. Also as expected, the average displacements are largest for BTO, with its deeper well and largest $\sigma$ and progressively smaller for STO and KTO. Both the O$_3$ and ABO$_3$ values have approximately the same BTO~:~STO~:~KTO displacement ratios of $\sim$5.5~:~$\sim$3~:~1.

\begin{table}[h!]
    \caption{\label{tab:displacements}%
    Various definitions of displacement and corresponding values (in \si{\AA}) for the materials BTO, STO and KTO (all numbers normalized to one formula unit and the center-of-mass motion).
    }
    \begin{tabular}{l r r r r}
    \toprule
        & \multicolumn{1}{c}{A} & \multicolumn{1}{c}{B} & \multicolumn{1}{c}{O$_3$} & \multicolumn{1}{c}{ABO$_3$}  \\
    \midrule
        BaTiO$_3$ &  0.007 &  0.114 &  0.139 &  0.107  \\
        SrTiO$_3$ &  0.016 &  0.044 &  0.073 &  0.056  \\
        KTaO$_3$  &  0.006 &  0.008 &  0.026 &  0.019  \\
    \bottomrule
    \end{tabular}
\end{table}

Six of the mass choices in Table~\ref{tab:masses} -- the O$_3$, ABO$_3$, reduced, AB/O$_3$. Wentzcovich and eigenvector masses -- give BTO~:~STO~:~KTO  displacement ratios close to the $\sim$5.5~:~$\sim$3~:~1 pattern identified for the O$_3$ and ABO$_3$ displacements of Table~\ref{tab:displacements}. Of these, the O$_3$ values give the most consistent match between the two definitions, with the displacements extracted from the ABO$_3$  mass choice underestimating the ABO$_3$ displacements in the double well, and all other mass choices providing values that are larger than any of the double-well displacement options. In particular, the tiny effective mass calculated from the phonon dispersion curvature, yields displacements comparable to the size of one unit cell which are likely unphysically large, even in the strongly fluctuating quantum regime. 

To summarize this section, if one is working with a model for polar quantum fluctuations that requires a separation into mass and displacement, then the best choice for the mass seems to be that of the lightest atoms. In addition to providing the best match with the calculations in mass-weighted coordinates, this choice is intuitive since the lightest atoms have the largest amplitude displacements in the soft polar mode. For perovskite titanates, the displacement of the B atom accidentally gives good results, since the mass of the titanium ion is equal to that of the three oxygens. It is not a good choice, however, for general non-titanate perovskites. This finding that the mass of the tunneling entity is best described by the oxygen mass implies that the mass shift on $^{16}$O~$\rightarrow$~$^{18}$O substitution should take the largest possible value of 12.5\%. Note, however, that this is still considerably smaller than the 100\% change on deuteration of ammonia.

\section{Conclusion}\label{sec:conclusion}

In summary, we find that a simple one-dimensional quantum model, incorporating only two parameters -- the energy difference between polar and non-polar structures and the frequency of the soft polar mode in the non-polar structure -- can straightforwardly distinguish between ferroelectric, quantum-paraelectric and paraelectric materials. We illustrate the model using conventional DFT and DFPT calculations for the cases of ferroelectric BaTiO$_3$, strongly quantum paraelectric SrTiO$_3$ and weakly quantum paraelectric KTaO$_3$. In addition, we provide a chart that indicates the behavior of any material given the values of these two parameters. 

Within our model, we find that experimentally accessible changes in lattice parameter have a strong effect on the shape of the double-well potential and hence the zero-point and low-lying energy levels, and readily transform STO and KTO from quantum paraelectric to ferroelectric behavior, consistent with experimental observations~\cite{BurkeStress1971,FujiiStressInduced1987,HaeniRoomtemperature2004,WordenweberInduced2007,VermaFerroelectric2015}. Complete isotopic substitution of oxygen-18 for oxygen-16, on the other hand, has a minimal effect, in contrast to the reported behavior in SrTiO$_3$~\cite{ItohFerroelectricity1999,ItohQuantum2000,RowleyFerroelectric2014,StuckyIsotope2016}. This could of course be a result of our model's simple description of the quantum mechanical behavior of the ions, although we note that the model accurately captures the change in tunneling frequency on deuteration of ammonia. Therefore, we suggest that care should be taken on interpreting changes in behavior after isotopic substitution in STO and KTO, and in particular that any possible changes in lattice parameters during the oxygen exchange process are accounted for. Finally, we summarize the various choices that have been made in the literature for phonon mass and use the results from our calculations in mass-weighted coordinates to determine the corresponding phonon displacement in each case. While we prefer the use of mass-weighted coordinates to avoid such an arbitrary and unphysical separation, we identify some choices which lead to reasonably sensible combinations of mass and displacement, particularly for BTO and STO.\\

\section{Acknowledgments}\label{sec:acknowledgments}
This work was funded by the European Research Council under the European Union’s Horizon 2020 Research and Innovation Program, grant agreement no.~810451. Computational resources were provided by ETH Zurich and the Swiss National Supercomputing Center (CSCS) under project~ID~s889.\\

\bibliography{PhD,Nicola}

\end{document}